\documentclass[pra,showpacs,twocolumn,floatfix,superscriptaddress]{revtex4}
\usepackage{times,amsmath,amsfonts,amssymb,latexsym}
\usepackage{graphicx,epsf}

\newcommand{\bra}[1]{\langle #1 |}
\newcommand{\ket}[1]{| #1 \rangle}

\newcommand{\be}{\begin{equation}}
\newcommand{\ee}{\end{equation}}
\newcommand{\ba}{\begin{eqnarray}}
\newcommand{\ea}{\end{eqnarray}}

\newcommand{\ignore}[1]{}
\def\CC{{\rm\kern.24em \vrule width.04em height1.46ex depth-.07ex
    \kern-.30em C}}
\def\P{{\rm I\kern-.25em P}}
\def\RR{{\rm
         \vrule width.04em height1.58ex depth-.0ex
         \kern-.04em R}}

\def\bbbc{{\mathchoice {\setbox0=\hbox{$\displaystyle\rm C$}\hbox{\hbox
to0pt{\kern0.4\wd0\vrule height0.9\ht0\hss}\box0}}
{\setbox0=\hbox{$\textstyle\rm C$}\hbox{\hbox
to0pt{\kern0.4\wd0\vrule height0.9\ht0\hss}\box0}}
{\setbox0=\hbox{$\scriptstyle\rm C$}\hbox{\hbox
to0pt{\kern0.4\wd0\vrule height0.9\ht0\hss}\box0}}
{\setbox0=\hbox{$\scriptscriptstyle\rm C$}\hbox{\hbox
to0pt{\kern0.4\wd0\vrule height0.9\ht0\hss}\box0}}}}
\def\bbbz{{\mathchoice {\hbox{$\sf\textstyle Z\kern-0.4em Z$}}
{\hbox{$\sf\textstyle Z\kern-0.4em Z$}}
{\hbox{$\sf\scriptstyle Z\kern-0.3em Z$}}
{\hbox{$\sf\scriptscriptstyle Z\kern-0.2em Z$}}}}

\begin{document}

\title{Decoherence and Entanglement Dynamics of Coupled Qubits}
\author{Gabriele Campagnano}
\affiliation{II Institut f\"{u}r Theoretische Physik, Universit\"{a}t Stuttgart, Germany}
\author{Alioscia Hamma}
\email{ahamma@perimeterinstitute.ca}
\affiliation{Perimeter Institute for Theoretical Physics\\
31 Caroline St. N, N2L 2Y5, Waterloo ON, Canada}
\affiliation{Massachusetts
Institute of Technology, Research Laboratory of Electronics\\
77 Massachusetts Ave. Cambridge MA 02139}
\author{Ulrich Weiss}
\affiliation{II Institut f\"{u}r Theoretische Physik, Universit\"{a}t Stuttgart, Germany}{}

\begin{abstract}
We study the entanglement dynamics and relaxation properties of a system of two interacting qubits in the two cases (I) two independent bosonic baths and (II) one common bath, at temperature $T$. The entanglement dynamics is studied in terms of the concurrence $\mathcal C (t)$ between the two spins and of the von Neumann entropy $S(t)$ with respect to the bath, as a function of time. We prove that the system does thermalize. In the case (II) of a single bath, the existence of a decoherence-free (DFS) subspace makes entanglement dynamics very rich. We show that  when the system is initially in a state with a component in the DFS the relaxation time is surprisingly long, showing the existence of {\em semi-decoherence free subspaces}. The equilibrium state in this case is not the Gibbs state. The entanglement dynamics for the single bath case is also studied as a function of temperature, coupling strength with the environment and strength of tunneling coupling. The case of the mixed state is finally shown and discussed.
\end{abstract}

\maketitle
\section{Introduction}
 The principle of superposition is the most important feature of quantum mechanics. It gives rise
to interference and to quantum entanglement. Among its endless implications, is the possibility of processing information
in a way that goes beyond the classical scheme of the Turing machine, with the fundamental consequence that information
is physical. Computer science is thus not a branch of pure mathematics. Moreover, quantum computation promises to
be able to solve certain computational problems with an exponential speed-up (for a review on quantum information, see \cite{qip}).

A quantum system will generically lose coherence when interacting with another quantum system. In this case, its evolution will not be unitary, and there will
be quantum correlations with the environment. The phenomenon of decoherence has been advocated as the solution of the measurement
problem in quantum mechanics, and the appearance of the classical world \cite{zurek, joos}. For quantum information, preserving
coherence is the most important and demanding problem in order to build a functioning quantum computer \cite{decoherence}. The
 loss of quantum behavior in a system can be measured by the loss of quantum correlations within the system, and the increase of
quantum correlations with the environment, or, in other words, by the entanglement within the system, and with the environment.
Unfortunately, there is not a general way to study entanglement between three parties. A relevant exception is constituted by the case
of two qubits because the \textit{concurrence} $\mathcal C$ between the two qubits is a valid measure of their entanglement even in a mixed state \cite{wot}.
Therefore, in the case of a composite quantum system constituted by two qubits and an environment, the entanglement between the qubits
can be characterized by their concurrence, while the von Neumann entropy $S$ can measure the entanglement between the system of the two
qubits and the environment.

In this paper we study a quantum system made of two interacting qubits coupled to a bosonic bath. We consider the following two cases:
 {\em (I)} for each spin one independent bath and {\em (II)} both spins coupled to the same bath.
 
The non equilibrium dynamics of the reduced density matrix $\rho(t)$ of the two qubits is studied within the Bloch-Redfield approach \cite{blum}. We assume that the environment is in thermal equilibrium at temperature $T$ and the system-bath coupling is weak. With this assumption, one obtains a set of coupled integro-differential equations for the elements of the reduced density matrix. In the Markov approximation, the equations of motion for $\rho(t)$ take then the form of simple linear differential equations that can be solved analytically.  Although quantum systems interacting with a bath are generically believed to relax, there are only a few examples where one can carry out all the calculations and prove directly how relaxation is achieved. In the case (I) we prove that the system always relaxes to the Gibbs state. The case (II) is more rich due to the existence of a decoherence-free subspace (DFS) \cite{dfs}. The novelty of our approach is also in taking in explicit consideration what happens if the system is initially in a state which has a component in the DFS, which has very important consequences. 

The main object of this work is the study of the quantities $\mathcal C(t)$ and $S(t)$ as a function of time and temperature $T$ for both the cases (I-II). Since we are interested in the formation of quantum correlations, we put emphasis in the case $T=0$ where the entropy $S$ has the meaning of entanglement with the bath degrees of freedom. The system of two spins in
bosonic bath has been studied in several papers \cite{braun,benatti,oh,zan1,wilhelm}. 

As mentioned above, the existence of the DFS reveals an important novel effect. There is an "interference" effect that decreases dramatically the decoherence rate when the initial state is in a coherent superposition between the DFS and its complement. We also see that the system relaxes to a state that is not the Gibbs state because the amplitude of the singlet state must stay constant. 
Moreover, we study entanglement dynamics and relaxation behavior for the system {\em (II)} as a function of temperature $T$ and coupling strength. 
Finally, we study the case of the mixed state, to show that the dramatic increase of relaxation time discussed above is genuinely a quantum effect.

\section{The Model}
 We consider a simple generalization of the {\em spin-boson}
problem \cite{leggett}, where two qubits
interact with each other via an Ising type coupling and are also coupled to a bosonic environment. The system Hamiltonian is
\be\label{hs}
H_S=-\frac{\Delta}{2}\sigma_x-\frac{\Delta}{2}\tau_x -\frac{v}{2} \sigma_z \tau_z .
\ee
We take $\hbar=k_B=1$.
Here  $\sigma$'s and $\tau$'s are the Pauli matrices on
the first and the second spin respectively. For simplicity bias terms are absent and the tunneling
coupling $\Delta$ is
the same for both spins.
The system Hamiltonian is trivially diagonalized (see appendix) and we will denote by $\{\ket{E_i}\}$ the basis of its eigenstates.
In case (I), the baths are modeled by the Hamiltonian
\be\label{bibath}
H_B^{(I)}=\sum_{\alpha,i=1,2}\omega_{\alpha,i}b_{\alpha,i}^\dag b_{\alpha,i}
\ee
The interaction Hamiltonian is
\be
H_{int}^{(I)}=\frac{1}{2}\sigma_z\sum_{\alpha}c_{\alpha,1}(b^\dag_{\alpha,1}+ b_{\alpha,1} )
+\frac{1}{2}\tau_z\sum_{\alpha}c_{\alpha,2}(b^\dag_{\alpha,2}+ b_{\alpha,2} )
\ee
Each spin is subject to one fluctuating force and the two forces are uncorrelated. The coefficients $c_{\alpha,i}$ give the strength of the
coupling of the system to each harmonic oscillator of the bath, which we assume to be weak.
For a gaussian model all the properties of the baths are described by the spectral density
$J_i(\omega)=\pi\sum_\alpha c^2_{\alpha,i}\delta(\omega_{\alpha,i}-\omega)$
which, in the case of ohmic baths that we consider here, takes the form
$J_i(\omega)=2\pi K_i \, \omega \exp(-\omega /\omega_c)$. Here $\omega_c$ is the cut-off frequency which is assumed to
be the largest energy scale in the problem.
We study the case $K_1=K_2\equiv\kappa/2\pi$,
with the two baths at the same temperature $T$.
In the case \textit{(II)} the same fluctuating force acts on both spins
so that $H_B$ has only one set  of harmonic oscillators.
The coupling of the spins to the reservoir in this second case  reads
\be
H_{int}^{(II)}=\frac{1}{2}(\sigma_z+\tau_z)\sum_{\alpha}c_{\alpha}(b^\dag_{\alpha}+ b_{\alpha} )
\ee
\section{Bloch-Redfield Approach}

 In this section we introduce the equations which allow us to
study the time evolution of the reduced density matrix of the two spins.
Within a Markov approximation, the matrix elements of the reduced
density matrix in the eigenvectors basis $\{\ket{E_i}\}$ obey the following generalized master equation \cite{blum}:
\be\label{master}
\dot{\rho}_{m',m}(t)=-i\omega_{m'm} \rho_{m'm}(t)+ R_{m'mn'n}\rho_{n'n}(t)
e^{i(\omega_{m'm}-\omega_{n'n})t}
\ee
where $R_{m'mn'n}$ is the Redfield tensor defined by $R_{m'mn'n}= -\sum_k (\delta_{mn}
\Gamma^+_{m'kkn'}+\delta_{n'm'}\Gamma^-_{nkkm}) +\Gamma^+_{nmm'n'}+\Gamma^-_{nmm'n'}$.
The rates  $\Gamma^\pm$ are given in the appendix. The indexes $m,m'n,n'$ run from $1$ to $4$. $E_1,...,E_4$ are the
eigenvalues of $H_S$ and $\omega_{mn}=E_m-E_n$.

Equation (\ref{master}) can be simplified by keeping only secular terms,
i.e. those terms such that the argument of the exponential is zero.
The equation for the diagonal elements ({\em populations}) of the density
matrix is
\begin{equation}\label{population}
\dot{\rho}_{mm}(t)=\sum_{m\ne n}\rho_{nn}(t)W_{mn}-\rho_{mm}(t)
\sum_{n \ne m}W_{nm},
\end{equation}
with $W_{mn}=\Gamma^+_{nmmn}+\Gamma^-_{nmmn}=2 {\mbox Re}\, \Gamma^+_{nmmn}$
For the  off diagonal elements ({\em coherences}) we have
\begin{equation}\label{coherences}
\dot{\rho}_{mn}(t)=(-i \omega_{mn}-\gamma_{mn})\rho_{mn}(t)+\sum_{kl}\Theta_{mnkl}
\, \rho_{kl}(t),
\end{equation}
with $\gamma_{m'm}=\sum_k\left(\Gamma^+_{m'kkm'}+\Gamma^-_{mkkm}\right)-
\Gamma^+_{mmm'm}-\Gamma^-_{mmm'm'}$ and $\Theta_{mnkl}=\Gamma^+_{lnmk}+\Gamma^-_{lnmk}$ if $\omega_{mn}=\omega_{kl}$ and
zero otherwise.
Due to the symmetries of the model the coefficients $\Gamma^{\pm}_{mmm'm'}$ are identically zero. To simplify the
calculation we disregard the Lamb shift to the eigenfrequencies due to
the imaginary part of the $\Gamma$'s. Within this approximation we
can write 
\begin{equation}\label{gamma}
\gamma_{mm'}=\frac{1}{2} \sum_k \left(W_{km'}+W_{km}\right).
\end{equation}

 We solve Eqs. (\ref{population}) and (\ref{coherences})
by Laplace transform and discuss separately the cases {\em (a-b)}.
\begin{figure}
  \includegraphics[width=9cm]{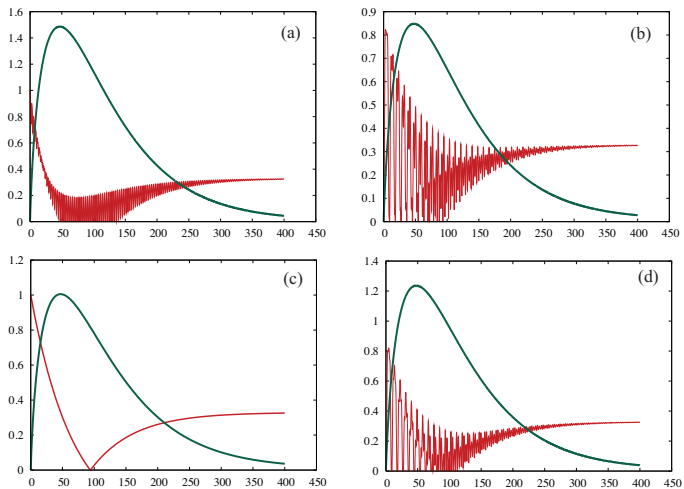}\\
  \caption{(Color online) Entanglement dynamics for the system with $\Delta=1,v=0.7$ for the case {\em (I)} of two independent baths at low temperature $\beta=10$. All the graphs show the Von Neumann entropy of the system of two spins and Concurrence between the two spins as a function of the time $t$. The initial states are $\ket{\Psi_a} =1/\sqrt{2}(\ket{\uparrow\downarrow}+\ket{\downarrow\uparrow})\in\mathcal H_\perp$ and $\ket{\Psi_b}=\ket{\uparrow\uparrow}\in\mathcal H_\perp$, $\ket{\Psi_c}=\ket{3}\in\mathcal H^{s}$, $\ket{\Psi_d}=\ket{\uparrow\downarrow}\in\mathcal H$, respectively. The two spins are initially disentangled with the environment. At $t=\infty$, the system always thermalizes in the ground state and therefore is a pure state: at short times the system gets quickly entangled with the environment due to fast decoherence, then it dissipates to reach the ground state. The relaxation time is of the order of $t\sim 400$. Notice the different behavior of the concurrence in the different cases. In particular, there are no oscillations for the case of the initial singlet state. When the system is very entangled with the environment, the concurrence dynamics changes qualitatively, before dissipation intervenes to damp its oscillations.}
  \label{bi}
\end{figure}

\begin{figure}
  \includegraphics[width=9cm]{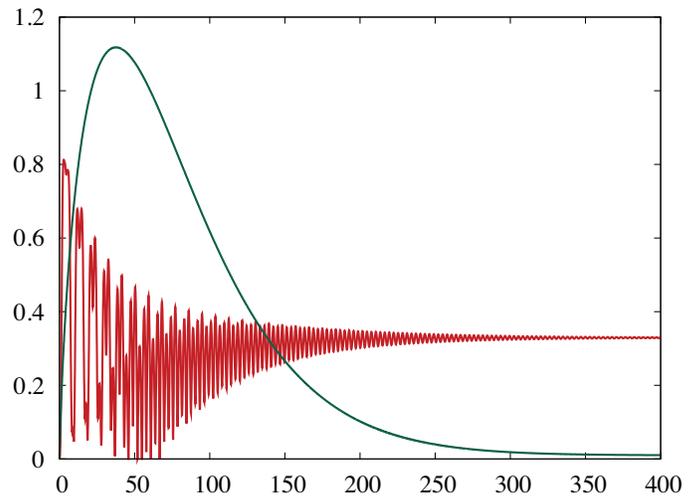}\\
  \caption{(Color online) Entanglement dynamics for the system with
  $\Delta=1,v=0.7$ for the case {\em (II)} of a single bath at low temperature
  $\beta=10$. All the graphs show the Von Neumann entropy of the system of two
  spins (blue) and Concurrence between the two spins (red) as a function of
  the time $t$. The initial state is $\ket{\Psi_a}
  =1/\sqrt{2}(\ket{\uparrow\downarrow}+\ket{\downarrow\uparrow})\in\mathcal
  H_\perp$. The two spins are initially disentangled with the bath. At
  $t=\infty$, the system thermalizes in the ground state and thus is
  disentangled again. At short times the system gets quickly entangled with
  the environment due to fast decoherence, then it dissipates to reach the
  ground state. Unlike the double bath case, there is an important
  relationship between entanglement with the bath and concurrence. The
  concurrence decreases, but when the system is very entangled with the
  bath, the structure of the oscillations changes qualitatively before the
  dissipation becomes relevant. The relaxation time is of the order of $t\sim
  400$.}
   \label{mono}
\end{figure}

\begin{figure}
  \includegraphics[width=9cm]{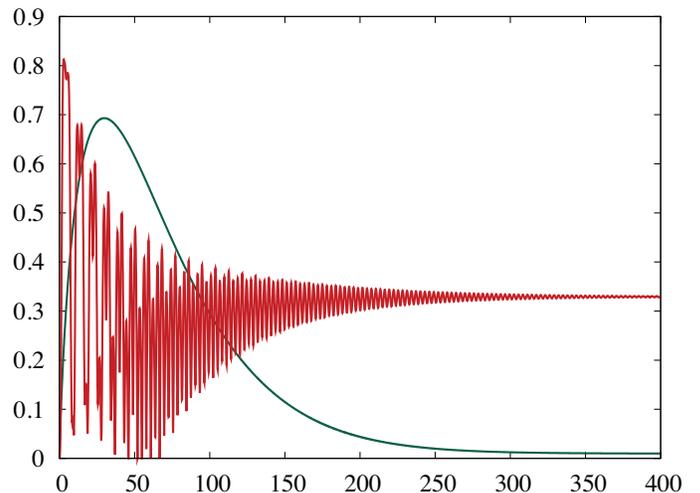}\\
  \caption{(Color online) Here we show the entanglement dynamics for the same value of parameters of the previous figure but for the initial state $\ket{\Psi_b}=\ket{\uparrow\uparrow}\in \mathcal
  H_\perp$. Since also this state does not have any component on the DFS $\mathcal H_s$, the qualitative features of $S(t),C(t)$ are the same than for the initial state  $\ket{\Psi_a}$ though the details of the evolution are different. The relaxation time is here $t\sim 400$ as well.}
  \label{monoB}
\end{figure}


\begin{figure}
  \includegraphics[width=9cm]{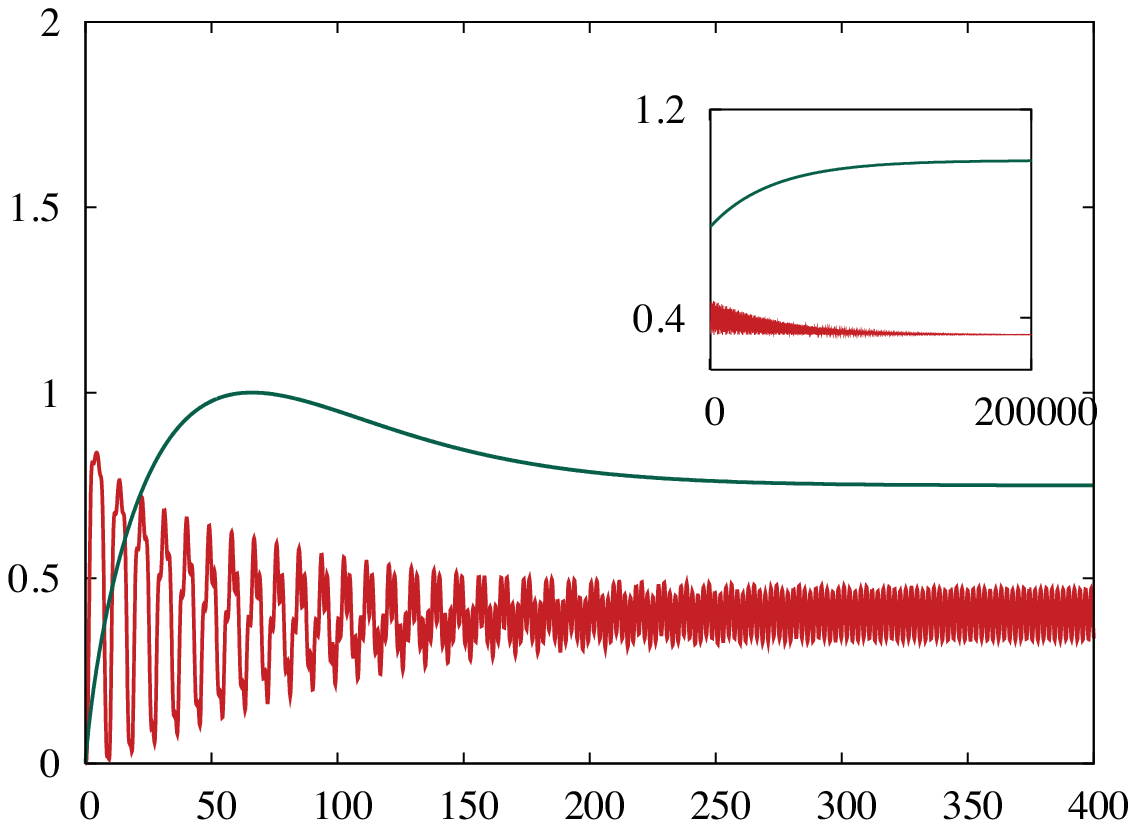}\\
  \caption{(Color online) Here we show the entanglement dynamics for the same system of Fig.2 but with initial state $\ket{\Psi_d} =\ket{\uparrow\downarrow}$. The component in $\mathcal H^s$ gives a dramatically longer relaxation time. The Von Neumann entropy will not equilibrate to zero so that there will be a residual significative entanglement with the bath. The concurrence oscillations are damping very slowly. In the inset, we show the behavior of $\mathcal C(t)$ and $S(t)$ at a very large $t$. The relaxation time is of the order of $t\sim4\times 10^5$. Moreover, the system-bath entanglement (Von Neumann entropy) reaches a minimum before slowing rising to the final value at the equilibrium.}
  \label{monolong}
\end{figure}

Direct calculation shows that all the coefficients which appear in
Eqs. (\ref{population}) and (\ref{coherences}) are not independent.
Before we proceed to brute force solution of the differential equations
it is convenient to employ all the symmetries of the problem.
From the spectrum of $H_S$ we obtain that  $\omega_{31}=\omega_{42}$
and $\omega_{43}=\omega_{21}$.  Moreover  $W_{mn}=\exp[-\beta \omega_{mn}]W_{nm}$.
For both cases {\em (I)} and {\em  (II)} the solution of  Eq.(\ref{coherences}) is straightforward once we notice that for this
model all the $\Theta$'s are zero. This simplification does not apply when the two spins have a different $\Delta$'s and one hads to take care of all term appearing in EQ>(\ref{coherences}).
We obtain
\begin{equation}\label{cohersol}
\rho_{ij}(t)=\rho_{ij}(0)\exp[- (\gamma_{ij} +i \omega_{ij})t]    \,\,\, \,\,\,\, i\ne j.
\end{equation}

In the case (I) direct calculation shows that
$W_{43}=W_{21}$ and $W_{42}=W_{31}$ and $W_{41}=W_{32}=0$.
In the case {\em (II)} the system is assumed to interact with a single ensemble of harmonic oscillators. It is well  known that in this case a DFS exists, {\em i.e.}, a subspace of the non interacting system which is  fully decoupled for the environment and where the time evolution is unitary \cite{dfs}.  For this simple model the DFS is one-dimensional and coincides with the singlet eigenstate $\ket{3}$.  It follows that
$W_{13},W_{23},W_{43}$ are all identically zero. Moreover also $W_{14}$ is zero as the
straightforward calculation shows.  Also in this case we are reduced with only two parameters $W_{42}$ and $W_{21}$ which in this case are not equal as in the double bath case. Of course the population $\rho_{33}$ of the singlet state remains constant in time.
In both cases (a-b), the analytic solution of Eq.(\ref{population}) is then obtained by taking the Laplace transform with respect to the time
and by solving a set of algebraic equations. The explicit expressions are reported in the appendix. We can write
the full solution of Eqs.(\ref{population}-\ref{coherences}) in a superoperator
form: $\rho (t) \equiv \mathcal E_t \rho(0)$.

\section{Entanglement dynamics and the DFS interference}

Having found the solution $\rho(t)$ for the time evolution of the density matrix of the system, we can proceed to the study of the entanglement dynamics. Let us call $\mathcal{H}^s$ the subspace of the singlet state $\ket{3}$ and $\mathcal{H}^\perp\equiv\mbox{span} \{\ket{1},\ket{2},\ket{4}\}$ the subspace orthogonal to it. We will initialize the system in the pure state $\rho(0)=\ket{\Psi} \bra{\Psi}$ corresponding to a generic superposition between the two subspaces; $\ket{\Psi} = A \ket{\phi_\perp} + B \ket{3}$, with $\ket{\phi_\perp}\in\mathcal H^\perp$.
Having obtained the solution $\rho(t)$ for the dynamics of the reduced system, we can study the von Neumann entropy $S(\rho(t))=-\mbox{Tr} (\rho \log\rho)$ and the two-spins concurrence $C(t)$ defined in \cite{wot}
\be
\mathcal{C}(\rho(t))\equiv\max(0,\sqrt{\lambda_1}-\sqrt{\lambda_2}-\sqrt{\lambda_3}-\sqrt{\lambda_4}), 
\ee
as a function of time. Here $\lambda_i$'s are the eigenvalues of $\rho(\sigma_{y}\otimes\sigma_{y})\rho^{*}(\sigma_{y}\otimes\sigma_{y}) $. We have performed a study in the space of the parameters $v,\Delta,\kappa,\beta=1/T$ and different initial states $\rho(0)$. The strength of the coupling $\kappa$ only changes the time scale of the evolution. The parameters $v,\Delta$ both set the time scale and a temperature scale. Moreover, for the case of the single bath they have an effect on the concurrence at the equilibrium. The effect of the temperature is that of making the system dissipating faster, and mixing the system.  Here, we want to focus on the important case of zero temperature because at $T=0$, the entropy $S(t)$ has the meaning of measuring the entanglement with the bath degrees of freedom. At $T>0$, it is not possible to distinguish the quantum correlations with the bath and the mixing due to the finite temperature. Initially the system is prepared in a pure state, and therefore $S(0)=0$. Then, the dynamics given by the interaction Hamiltonian will entangle the system with the environment. Let us examine now the two cases.

{(I) Two independent baths.---} In Fig.\ref{bi} are plotted $\mathcal C(t)$ and $S(t)$ for the four initial states $\ket{\Psi_a} =1/\sqrt{2}(\ket{\uparrow\downarrow}+\ket{\downarrow\uparrow}), \ket{\Psi_b}=\ket{\uparrow\uparrow}$ in the irreducible subspace $\mathcal H_\perp$, the singlet state $\ket{\Psi_c}=\ket{3}\equiv 1/\sqrt{2}(\ket{\uparrow\downarrow}-\ket{\downarrow\uparrow})$, and $\ket{\Psi_d}=\ket{\uparrow\downarrow}$. The choice of parameters is $v=0.7,\Delta=1,\kappa=0.01$ and $\beta=10$ that is a very low temperature for the system. The concurrence $\mathcal C(t)$ goes to zero as the system decoheres. Since the time of decoherence is much faster than dissipation, most of the concurrence is lost at earlier times, then it degrades more slowly. If the system is initialized in $\mathcal H_\perp$, it will be oscillating with damping oscillations. If it is initialized in the singlet state, it just leaks towards the other states (in particular, the ground state) without oscillations because $\ket{3}$ is an eigenstate of $H_S$ (Fig.\ref{bi}c). In all cases the concurrence revives after having gone to zero, in the point where the entanglement with the environment is maximum, a sign of monogamy of entanglement. An important effect is that, when the system gets more entangled with the baths, the concurrence dynamics changes qualitatively and the some oscillations increase their amplitude (see, in particular, Fig.\ref{bi}b).

We can prove that the system relaxes taking the limit   $\rho^{(bi)}_{eq}\equiv\lim_{t\rightarrow \infty}\rho^{(bi)}(t)$, and obtain, for a generic $\beta$, $
\rho^{(bi)}_{eq} =Z^{-1}\mbox{diag}(e^{-\beta E_1} ,e^{-\beta E_2} ,e^{-\beta E_3} ,e^{-\beta E_4} )$ where $Z= \sum_{i=1}^4 e^{-\beta E_i}$ is the partition function of the system. The system thermalizes in the Gibbs state. This is what everyone would expect, a system is supposed to thermalize in the Gibbs state. It is though always very difficult to prove relaxation to equilibrium in  concrete examples and this is one of our results. At $T=0$ the Gibbs state is the ground state and therefore the asymptotic value for $S$ is zero. At $T>0$, the von Neumann entropy will first increase to a high value due to the fast decoherence, and then, when dissipation kicks in, slowly decrease to its asymptotic value given by the entropy of the Gibbs state. Notice that the relaxation time is of the order of $t\sim 400$, independently of the initial state.

{(II) Single bath.---} The behavior for the system in a single bath is completely different. In this case, the singlet subspace $\mathcal H^{s}$ is a DFS \cite{dfs}. If the system is prepared in the singlet state $\rho(0)=\ket{3}\bra{3}$, it will stay there forever, so the case $A=0$ is trivial. If we prepare the state in the subspace $\mathcal H_\perp$, there can be no effect due to the DFS. Nevertheless, the entanglement dynamics is very interesting. In Fig.\ref{mono}a, the concurrence $\mathcal C(t)$ and the von Neumann entropy $S(t)$ are plotted for the initial states $\ket{\Psi_a} =1/\sqrt{2}(\ket{\uparrow\downarrow}+\ket{\downarrow\uparrow})\in\mathcal H_\perp$. At the beginning, the system is not entangled with the bath, and the concurrence starts decreasing, due to fast decoherence, in a fashion very similar to the case of the double bath. Then notice, that also here, when the system is very entangled with the bath, the concurrence dynamics changes again qualitatively. When the dissipation becomes relevant, the oscillations damp and the system relaxes to the Gibbs state. At the plotted temperature $\beta=10$, this is practically the ground state so that the entanglement with the bath is zero. The equilibration time is of the same order of magnitude than the case {\em (I)} with the two independent baths. In Fig.\ref{mono}b the initial state is $\ket{\Psi_d}=\ket{\uparrow\downarrow}$, which is the equal superposition between $\ket{\Psi_a}$ and the singlet state $\ket{\Psi_c}\equiv\ket{3}$. Now things change dramatically. The concurrence dynamics is qualitatively completely different, and it damps in an extremely slower way. In Fig.\ref{monolong} is shown the behavior at large times: the equilibration time is of the order of $t\sim 4\times 10^5$, three orders of magnitude more than the usual. How is it possible? Why the oscillations of the system are damped in such a slow way? Even if "half" of the system is in the DFS, it could still be that the decoherence time will depend only on the part that is in $\mathcal H_\perp$. This is exactly the case if we prepare the initial state in the mixed state $\rho(0)=1/2(\ket{\Psi_1}\bra{\Psi_1}+\ket{3}\bra{3})$. The evolution equations $\rho(t)=\mathcal E_t \rho$ are linear and we would expect the same type of oscillations than for the first term only, since the second one is constant in time. But that is not the case. The state is prepared in a {\em coherent} superposition of $\ket{\Psi_1}$ and the singlet state $\ket{3}$, so this means that the off-diagonal terms in $\rho(0)$ show up in $\rho(t)= \mathcal E_t \rho(0)$ in a way that makes the decoherence much slower. We say it is an \textit{interference} effect because it discriminates between a coherent superposition and a classical mixture. As we shall see in section \ref{mixedstate}, there is no such effect if the initial state $\rho(0)$ is a classical mixture of states in $\mathcal H^s$ and $\mathcal H^\perp$. 

In order to understand the long decay time when the system is initialized in a state with a component in 
the DFS  $\mathcal H^s$, let us look at the decay of $\rho_{13}$ and compare it to the one of, say, $\rho_{12}$. The oscillation behaviors are comparable but the exponential decays are very different. 
These decays are governed by the rates $\gamma_{13}$ and $\gamma_{12}$, as one can see from  Eq.(\ref{cohersol}).
Using Eq.(\ref{gamma}) and the fact the $W_{13}$ is zero we readily find that at low temperature    $\gamma_{13} \propto \coth(\beta \omega_{21}/2)-1$  while $\gamma_{12} \propto \coth(\beta \omega_{21}/2)+1$. With our choice of the parameters we have $\gamma_{12}/\gamma_{13}\sim 10^3$
as illustrated in the plots.

Not only the system thermalizes with a longer time scale, but it does not relax to the Gibbs state. After all, we expect the population of the singlet state to be a constant of the motion. Taking in $\rho^{(mono)}(t)$ the limit for $t\rightarrow\infty$, we find
\be
\rho_{eq}^{(mono)}=\mbox{diag} \left(\frac{|A|^2}{Z_3}e^{-\beta E_1},\frac{|A|^2}{Z_3}e^{-\beta E_2},|B|^2,\frac{|A|^2}{Z_3}e^{-\beta E_4}\right)
\label{eqmono}
\ee
\ignore{\be
\rho_{eq}^{(mono)} =
\left(
\begin{array}{cccc}
\frac{|A|^2}{Z_3}e^{-\beta E_1} & 0 & 0 & 0 \\
0 &\frac{|A|^2}{Z_3} e^{-\beta E_2}& 0 & 0 \\
0 & 0  & |B|^2 & 0\\
0 & 0 & 0 & \frac{|A|^2}{Z_3}e^{-\beta E_4}
\end{array}\label{eqmono}
\right)
\ee}
where $Z_3=\sum_{i\ne 3} e^{-\beta E_i}$ is the partition function over the irreducible subspace $\mathcal{H}^\perp$.
Because of the constant term $\rho_{33}^{(mono)}=|B|^2$, the entanglement with the bath can be non vanishing even at the equilibrium at zero temperature. For instance, at $T=0$ the equilibrium value for the Von Neumann entropy $S$ is given by $S= |A|^2\log |A|^2+ (1-|A|^2)\log (1-|A|^2)$ and is obviously maximized by $|A|^2=1/2$ for which we have $S=1$, as it was also argued in \cite{zan1}. The concurrence $C$ at zero temperature still depends also on $v,\Delta$. For the case study of $v=0.7,\Delta=1$ the equilibrium concurrence at $T=0$ is $C_{eq}\simeq 0.33$.

\section{Study in temperature and coupling strengths}
In this section we study the behavior of the entanglement dynamics in the system {\em (II)} as a function of the coupling strength with the environment $\kappa$, the temperature $\beta$, and the parameter $\Delta$. The initial state is $\ket{\Psi_a} =1/\sqrt{2}(\ket{\uparrow\downarrow}+\ket{\downarrow\uparrow})$.

\begin{figure}
  \includegraphics[width=8cm]{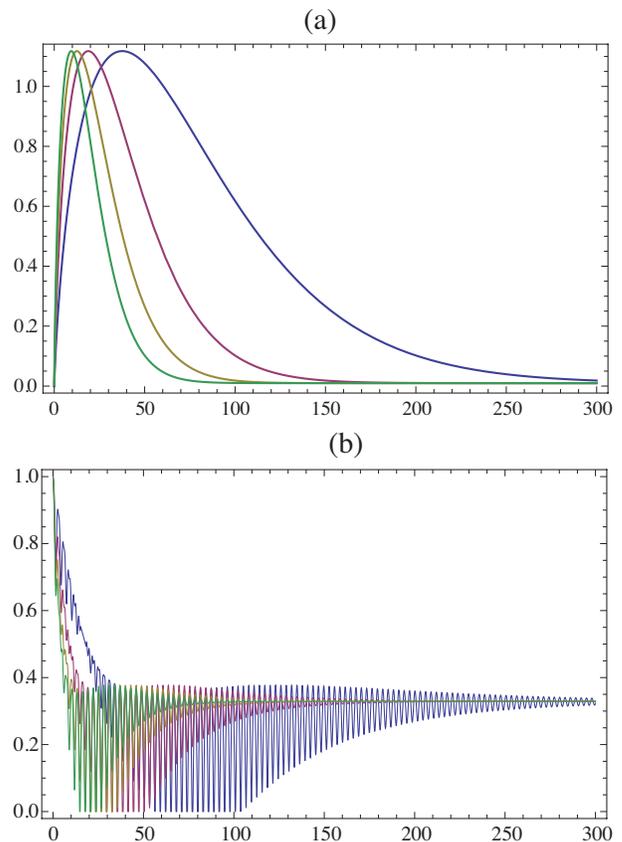}
  \caption{(Color online) Von Neumann entropy $S(t)$ (a) and Concurrence $C(t)$ (b) as a function of time for different values of the coupling with the environment $\kappa = 0.1,0.2,0.3,0.4$, for the initial state is $\ket{\Psi_a} =1/\sqrt{2}(\ket{\uparrow\downarrow}+\ket{\downarrow\uparrow})$. The temperature is $\beta=10$ and $v=0.7, \Delta=1$. A stronger coupling has the effect of shrinking the time scale for the entanglement dynamics without any other qualitative changes.}
  \label{kstudy}
\end{figure}

The results of the study in $\kappa$ are plotted in Fig.(\ref{kstudy}), for a low temperature $\beta=10$. We expect that a greater coupling with the environment will not make qualitative changes, as long as the hypotheses of weak coupling are still satisfied. We expect that the time scale of the system will shrink for larger couplings. Fig.(\ref{kstudy}) confirms this physical insight in the behavior of both $C(t)$ and $S(t)$ at every time scale.

\begin{figure}
  \includegraphics[width=8.5cm]{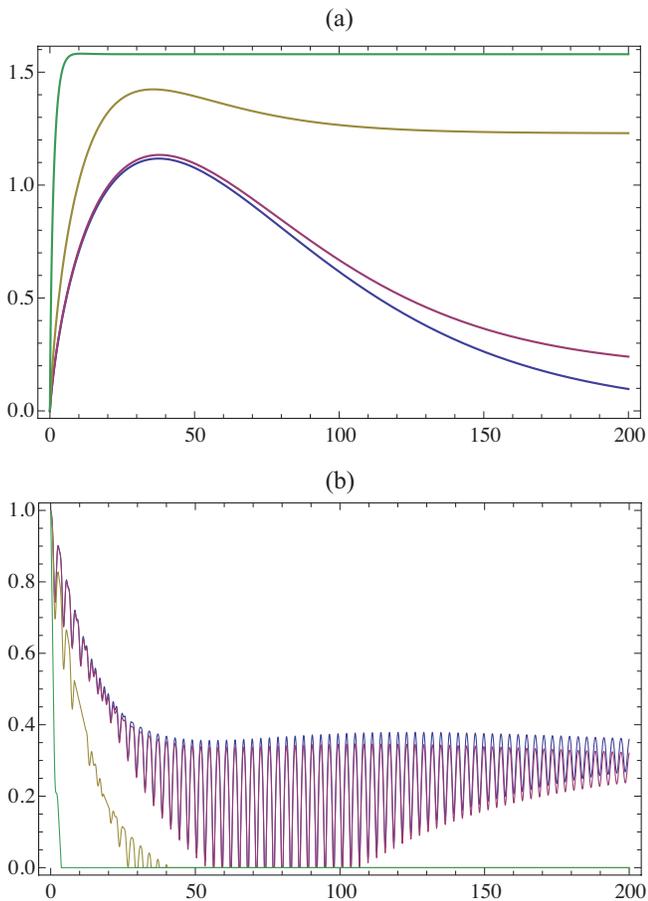}
  \caption{(Color online) Study in temperature of the entanglement dynamics for the case (II) of a single common bath. Here $v=0.7, \Delta=1,\kappa = 0.01$. The initial state is $\ket{\Psi_a} =1/\sqrt{2}(\ket{\uparrow\downarrow}+\ket{\downarrow\uparrow})$. In the graph (a) is plotted the Von Neumann entropy $S(t)$ as a function of time for different values of the temperature $\beta = 20,5,1,0.1$.  At low temperatures the state approaches the ground state as equilibrium state. The non monotonic behavior of $S(t)$ shows the different time scales of decoherence and dissipation. In the graph (b) we show $C(t)$ for the same temperatures. At low temperatures the phenomenon of entanglement revival occurs.}
  \label{tempstudy}
\end{figure}

The second study is the behavior of the system in temperature. In Fig.(\ref{tempstudy}) we have plotted the time evolution of $S(t)$ and $C(t)$ for different temperatures $\beta$. At high temperature $\beta=.1$, the system decoheres very rapidly and the entanglement dynamics is trivial: it entangles and mixes with the environment and the two spins disentangle from each other. At a medium temperature $\beta=1$, the process of entanglement and dissipation towards the environment is smoother, and the concurrence dynamics is less trivial, eventually though, the two spins disentangle from each other. At low temperatures $\beta= 5,20$, the system shows the most interesting behavior. Now the Von Neumann entropy can be interpreted as just the entanglement with the environment. Its rising and decay marks the two phases of decoherence and dissipation. Initially, the system rapidly decoheres by entangling with the environment. The time scale of decoherence is much smaller than the one of dissipation. Then, dissipation intervenes, and at very low temperature the system must fall into the ground state, which is a pure state and thus disentangled with the environment. 

The behavior of Concurrence $C(t)$ is the most interesting as a function of temperature. At high temperatures, the concurrence is rapidly damped down because of the entanglement with the environment. The entanglement is monogamous so as the system entangles with the environment, the mutual entanglement between the spins decreases. Once zero, even though the system gets more and more mixed, the Concurrence remains zero. At high temperature there is no quantum correlation left in the system. At low temperatures instead, the Concurrence revives and its behavior is non monotonic. The Concurrence equilibrates to a non-zero value at 

 with damped oscillations. The amplitude of the oscillations changes pattern when the system is very entangled with the environment.

\section{The mixed state case}\label{mixedstate}
\begin{figure}
  \includegraphics[width=8.5cm]{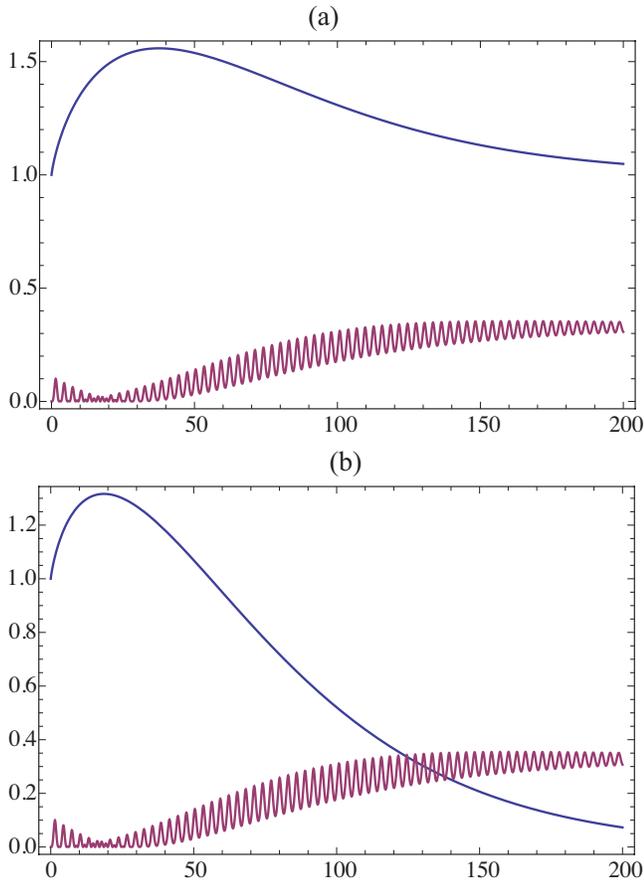}
  \caption{(Color online)  $S(t)$ and $C(t)$, for low temperature $\beta=20$. The values for the other parameters are $v=0.7,\Delta=1,\kappa = 0.01$. The initial states are (a) the classical mixture $\rho_{mix}^{(1)}$ of two states belonging to $\mathcal H_s$ and $\mathcal H^\perp$. In (b) we see the entanglement dynamics for the state $\rho_{mix}^{(2)}$ which is a classical mixture of two states belonging both to $\mathcal H^\perp$. The relaxation times are similar. Both states show revival in the Concurrence. The final entanglement with the bath is different in the two cases because $\rho_{mix}$ cannot be the ground state since in the mixture the component on  $\mathcal H_s$ is constant while $\rho_{mix}^2$ goes to the ground state.}
  \label{mixedstudy}
\end{figure}

In this section, we study the behavior of the system prepared in a initial classical mixed state. The common bath is at the extremely low temperature $\beta =20$.  Let us define the density matrices  $\rho_a = \ket{\Psi_a}\bra{\Psi_a}=(\ket{\uparrow\downarrow}+\ket{\downarrow\uparrow})(\bra{\uparrow\downarrow}+\bra{\downarrow\uparrow})/2$, $\rho_{b} = \ket{\Psi_b}\bra{\Psi_b}=\ket{\uparrow\uparrow}\bra{\uparrow\uparrow}$, $\rho_c = \ket{\Psi_c}\bra{\Psi_c}=\ket{3}\bra{3}= (\ket{\uparrow\downarrow}-\ket{\downarrow\uparrow})(\bra{\uparrow\downarrow}-\bra{\downarrow\uparrow})/2$. We want to show that if we prepare the system in the initial classical mixture $\rho_{mix}^{(1)} = (\rho_a + \rho_c)/2$, there is no trace of the behavior obtained when we have a coherent superposition of a state in the DFS $\mathcal H_s$ with one in $\mathcal H^\perp$. We will compare the entanglement dynamics with the initial preparation of the classical mixture of two states in $\mathcal H^\perp$, namely  $\rho_{mix}^{(2)} = (\rho_a + \rho_b)/2$. The plot of Fig.\ref{mixedstudy} shows the results. Let us first look at the evolution for the entanglement $S(t)$ with the bath. The state $\rho_{mix}^{(1)}$ gets very entangled with the bath during the fast decoherence period. Then when dissipation becomes important, the entanglement decreases. Nevertheless, the equilibrium state cannot be the Gibbs state because the population in the DFS $\mathcal H^s$ is constant. Therefore the final state is not the ground state for the system and some mixture is present. This mixture does not mean that there is residual entanglement with the bath even though we are at extremely low temperature. The state $\rho_{mix}^{(2)}$ instead, dissipates towards the ground state because there is no initial population in the initial state and it is, therefore, a pure state at the equilibrium, disentangled with respect to the bath. In both cases, the relaxation times are comparable $t\sim 400$ like in the case of Figs.\ref{mono}-\ref{monoB}. The concurrence $C(t)$ shows a similar pattern. In both cases we have revival of the concurrence after it hits zero and similar equilibration values. Again, the graphs show that the relaxation time is about $t\sim 400$.

This study shows that the dramatic increase in the relaxation time shown in Fig.\ref{monolong} is due to the quantum superposition of state in $\mathcal H_s$ and $\mathcal H^\perp$. It is a purely quantum effect. This means that if we consider for instance the subspace $\mathcal H^{semi}\equiv \mbox{span} \{\ket{3}, \ket{\Psi_a}\}$, although it is not a DFS, the relaxation time for states in this subspace is much larger than that of states its orthogonal complement, and this is because $\mathcal H^{semi}$ contains a DFS. We call $\mathcal H^{semi}$ a {\em semi decoherence-free subspace}. The existence of such subspaces is important in quantum computation because one can protect quantum memory and quantum information processing even in absence of a real DFS, which is often the case in presence of perturbations. Moreover, it enlarges the dimension of the subspace in which the information is protected. In the case studied for instance, the DFS is trivial because is one-dimensional, and no information (classical or quantum) can be encoded. Nevertheless, the subspace $\mathcal H^{semi}$ is two dimensional and one can encode a qubit in it. So an array of pairs of spins $1/2$ could constitute a good quantum register, of course for the model of noise presented here.

\section{Conclusions and Outlook}

In this article, we have thoroughly studied the system of two interacting qubits in a bosonic environment, in both the cases where each spin is interacting with its own bath, and where they share a common bath. We solved the master equation for the reduced system of the two qubits and studied the entanglement dynamics by means of Concurrence and von Neumann entropy. In the case of the single bath, the entanglement dynamics is very rich. The concurrence oscillates violently in a time window when the system is very entangled with the environment but the dissipative effects are not yet important. If the state is initialized in a coherent superposition with a component in a decoherence-free subspace, there is an interference effect that changes the decoherence rate and the entanglement dynamics. This opens the interesting possibility of doing quantum computation "straddling" a DFS, thus having a bigger code for computation, that still has much better protection than decoherence than the ones that are orthogonal to the DFS. Moreover, conditions for the existence of DFS are considered unstable. Our results show that an approximate notion of decoherence-free subspace is possible and useful for protecting quantum information.

We also proved that this system thermalizes, but not always in the Gibbs state. Upon the completion of this work, we noticed the study of \cite{petruccione}, showing, for a different model of two qubits in two baths, similar results for the concurrence dynamics.

\textit{Acknowledgements.---} The authors acknowledge discussion with D.~Lidar, S.~Lloyd and P.~Zanardi. Research at Perimeter Institute for Theoretical Physics is supported in part by the Government of Canada through NSERC and by the Province of Ontario through MRI. This project was partially supported by a grant from the Foundational Questions Institute (fqxi.org), a grant from xQIT at MIT. Financial support by the DFG through SFB/TR 21 is gratefully acknowledged.

\appendix
\section*{Appendix}
\subsection{Eigenvectors and eigenvalues of $H_S$}

We take as basis the eigenvectors of $\sigma_z \tau_z$:  
$|++\rangle$, $|+-\rangle$, $|-+\rangle$,
$|--\rangle$. In this basis $H_S$ simply  reads

\[
H_S=-\frac{1}{2}
\left[
\begin{array}{cccc}
v & \Delta & \Delta & 0 \\
\Delta & -v & 0 & \Delta \\
\Delta & 0  & -v & \Delta \\
0 & \Delta & \Delta & v
\end{array}
\right]
\]
We have the following  eigenenergies
\[
E_1=-\frac{1}{2}\sqrt{v^2+4 \Delta^2} \,\,\, \,\,\,
E_2=-\frac{v}{2}
\]
\[
E_3=\frac{v}{2} \,\,\,\,\,\,
E_4=\frac{1}{2}\sqrt{v^2+4 \Delta^2}
\]
The corresponding eigenvectors are

\[
|1\rangle=[r_+,s_+,s_+,r_+],
\]
\[
|2\rangle=\frac{1}{\sqrt{2}}[-1,0,0,1],
\]
\[
|3\rangle=\frac{1}{\sqrt{2}}[0,-1,1,0],
\]
\[
|4\rangle=[r_-,s_-,s_-,r_-].
\]

For notation simplicity we have introduced

\[
r_{\pm}=\frac{1}{2}\sqrt{1\pm\frac{v}{\sqrt{v^2+4\Delta^2}}},
\]
and
\[
s_{\pm}=\pm \Delta [4\Delta^2+v(v\pm\sqrt{v^2+4\Delta^2})]^{-\frac{1}{2}}.
\]

\subsection{Matrix elements of the reduced density matrix}

We present here the explicit calculation  of the 
rates appearing in the Master equation 
\ref{master} that we use to solve for the 
dynamics of the reduced density matrix $\rho_S$.
As explained in the main text  in the 
case of one single bath for both spins a (one dimensional) DFS 
exists. Because of the absence of bias terms in the 
Hamiltonian $H_S$ this  DFS happens to coincide with the
eigenvector $|3\rangle$.
 
\textit{Uncorrelated baths----} Assuming no correlation between 
the two baths,according to \cite{blum} we have

\begin{eqnarray}\label{gamma+}
\Gamma^+_{mkln}=\frac{1}{4}\sum_{i=1,2}\langle m|s_z^i|k \rangle
\langle l|s_z^i|n \rangle\\
\times\int_0^\infty dt \exp(-i \omega_{ln} t)
\langle X_i(t)X_i \rangle_\beta,
\end{eqnarray}
and

\begin{eqnarray}
\Gamma^-_{mkln}=\frac{1}{4}\sum_{i=1,2}\langle m|s_z^i|k \rangle
\langle l|s_z^i|n \rangle\\
\times\int_0^\infty dt \exp(-i \omega_{mk} t )
\langle X_i X_i(t) \rangle_\beta.
\end{eqnarray}
Where $s_z^1=\sigma_z$ and $s_z^2=\tau_z$, and $X_i=\sum_\alpha c_{\alpha,i}
(b^\dag_{\alpha,i}+b_{\alpha,i})$, we use the interaction representation
$X_i(t)=\exp(-i H_B t) X_i \exp(i H_B t)$. Here $\langle \cdot \rangle_\beta$ 
is the thermal equilibrium average over the bath's degrees of freedom. 

A direct calculation shows that

\[
 \langle X_i(t) X_i \rangle=\int_0^{\infty} d\omega J_i(\omega)
\left[ \coth(\frac{\beta \omega}{2})\cos(\omega t)-i\, \sin(\omega t)
 \right].
 \]

The only quantity we need to compute is an integral of the form:
 \begin{multline*}
 I_i^{\pm}(\omega)=\int_0^\infty dt \, e^{-i \omega t}
 \int_0^{\infty} d\omega' J_i(\omega') \\ \times
\left[ \coth(\frac{\beta \omega'}{2})\cos(\omega' t) \mp
i \, \sin(\omega' t)
 \right]
 \end{multline*}

 For $\omega > 0$  we obtain
 (the case $\omega <0$ is obtained by complex conjugation):
 \begin{multline*}
 I_i^{\pm}(\omega)=\frac{\pi}{2}J_i(\omega)
\left[ \coth \frac{\beta \omega}{2} \mp 1\right]
 \\+
 i {\cal P}\int_0^\infty d \omega' J_i(\omega')
 \left[\frac{\omega}{\omega'^2-\omega^2}\coth \frac{\beta \omega'}{2}
\mp \frac{\omega'}{\omega'^2-\omega^2}
\right],
 \end{multline*}
 ${\cal P}$ indicates the Cauchy principal value.
The imaginary  part  can be absorbed as shift of the
free oscillation frequencies of the system.
For simplicity we disregard it.
Because of the symmetries of $H_S$
we only need to calculate  $W_{21}$ and $W_{31}$,
we obtain

\[
W_{31}=\frac{\pi \Delta^2 k}{\sqrt{v^2+\Delta^2}}
\left[\coth(\frac{\beta \omega_{31}}{2})-1 \right],
\]
and
\[
W_{21}=\frac{\pi \Delta^2 k}{\sqrt{v^2+\Delta^2}}
\left[\coth(\frac{\beta \omega_{21}}{2})-1 \right].
\]

We report here the explicit solution of the diagonal elements of 
the density matrix in the diagonal basis.
We introduce $\Gamma_1=W_{21}+W_{12}$ and  
$\Gamma_2=W_{31}+W_{13}$, using this notation we have

\begin{widetext}

\begin{small}
\begin{multline*}
\rho_{11}(t)=
\left[{\left(1+e^{\beta  \omega _{21}}\right)
   \left(1+e^{\beta  \omega _{31}}\right)}\right]^{-1}
e^{-t \left(\Gamma _1+\Gamma _2\right)} \left[\left(e^{t \Gamma
   _2+\beta  \omega _{31}} \left(1+e^{\beta  \omega _{21}}\right)+e^{t
   \Gamma _1+\beta  \omega _{21}}-e^{\beta  \left(\omega _{21}+\omega
   _{31}\right)}+e^{t \Gamma _1+\beta  \left(\omega _{21}+\omega
   _{31}\right)}+1\right) \rho _{11}(0)
   \right. \\  \left.
   +e^{\beta  \omega _{21}} \left(-1+e^{t
   \Gamma _1}\right) \left(1+e^{\beta  \omega _{31}}\right)
   \rho_{22}(0)+e^{\beta  \omega _{31}} \left(-1+e^{t \Gamma _2}\right)
   \left(e^{\beta  \omega _{21}} \left(-1+e^{t \Gamma
   _1}\right)+\left(1+e^{\beta  \omega _{21}}\right) \rho
   _{33}(0)\right)\right],
\end{multline*}

\begin{multline*}
\rho_{22}(t)=\left[{\left(1+e^{\beta  \omega _{21}}\right)
   \left(1+e^{\beta  \omega _{31}}\right)}\right]^{-1}
e^{-t \left(\Gamma _1+\Gamma _2\right)}\left[\left(-e^{t \Gamma _2+\beta
   \omega _{31}} \left(1+e^{\beta  \omega _{21}}\right)+e^{t \Gamma
   _1}+e^{\beta  \left(\omega _{21}+\omega _{31}\right)}+e^{t \Gamma
   _1+\beta  \omega _{31}}-1\right) \rho _{11}(0)
      \right. \\  \left.
   +\left(e^{t \Gamma
   _1}+e^{\beta  \omega _{21}}\right) \left(1+e^{\beta  \omega
   _{31}}\right) \rho _{22}(0)+e^{\beta  \omega _{31}} \left(-1+e^{t \Gamma
   _2}\right) \left(-\left(1+e^{\beta  \omega _{21}}\right) \rho
   _{33}(0)+e^{t \Gamma _1}+e^{\beta  \omega _{21}}\right)\right],
\end{multline*}

\begin{multline*}
\rho_{33}(t)=
\left[{\left(1+e^{\beta  \omega _{21}}\right)
   \left(1+e^{\beta  \omega _{31}}\right)}\right]^{-1}
e^{-t \left(\Gamma _1+\Gamma _2\right)}\left[\left(-e^{\beta
   \left(\omega _{21}+\omega _{31}\right)} \left(-1+e^{t \Gamma
   _1}\right)+e^{t \Gamma _2}-e^{t \Gamma _1+\beta  \omega _{21}}+e^{t
   \Gamma _2+\beta  \omega _{21}}-1\right) \rho _{11}(0)
      \right. \\  \left.
  -e^{\beta  \omega
   _{21}} \left(-1+e^{t \Gamma _1}\right) \left(1+e^{\beta  \omega
   _{31}}\right) \rho_{22}(0)+\left(e^{t \Gamma _2}+e^{\beta  \omega
   _{31}}\right) \left(e^{\beta  \omega _{21}} \left(-1+e^{t \Gamma
   _1}\right)+\left(1+e^{\beta  \omega _{21}}\right) \rho
   _{33}(0)\right)\right].
\end{multline*}

\end{small}
\end{widetext}

The calculation of the off-diagonal elements does not present any difficulties 
and it obeys the relation given in the main text. 
The explicit expressions of the 
$\gamma$'s are directly related to  
$W_{21}$ and $W_{31}$ via Eq.\ref{gamma}.

\textit{Single bath----} In the case of one single bath we have

\begin{multline*}
\Gamma^+_{mkln}=\frac{1}{4} \langle m|\sigma_z+\tau_z |k \rangle
\langle l|\sigma_z +\tau_z |n \rangle \\ \times
\int_0^\infty dt \exp(-i \omega_{ln} t)
\langle X(t)X \rangle_\beta,
\end{multline*}
and
\begin{multline*}
\Gamma^-_{mkln}=\frac{1}{4}\langle m|\sigma_z+\tau_z |k \rangle
\langle l|\sigma_z+\tau_z |n \rangle \\ \times
\int_0^\infty dt \exp(-i \omega_{mk} t )
\langle X X(t) \rangle_\beta.
\end{multline*}

As anticipated the state $|3\rangle$ is totally 
uncoupled to the other states,
i.e. $W_{31}=W_{32}=W_{34}=0$ as the direct calculation shows.
In this case $\rho_{33}(t)=\rho_{33}(0)$. We have

\[
W_{42}=\frac{2 \pi \Delta^2 k}{\sqrt{v^2+\Delta^2}}
\left[\coth(\frac{\beta \omega_{42}}{2})-1 \right],
\]
and
\[
W_{21}=\frac{2 \pi \Delta^2 k}{\sqrt{v^2+\Delta^2}}
\left[\coth(\frac{\beta \omega_{21}}{2})-1 \right].
\]

The two  coupled equations for $\rho_{11}$ and $\rho_{22}$
given by Eq. \ref{population} are solved by Laplace transform,
we  obtain 

\begin{widetext}
\[
\rho_{11}(\lambda)=\frac{\lambda  \left(\lambda +W_{12}+W_{24}+W_{42}\right) \rho _{11}(0)+W_{12} \left(\lambda
   \rho _{22}(0)-W_{24} \left(\rho _{33}(0)-1\right)\right)}{\lambda  \left(W_{12} \left(\lambda
   +W_{24}\right)+\left(\lambda +W_{21}\right) \left(\lambda +W_{24}+W_{42}\right)\right)},
\]

\[
\rho_{22}(\lambda)=\frac{W_{21}(W_{24}+\lambda (\rho_{11}(0)+\rho_{22}(0))
-W_{24} \lambda \rho_{33}(0)) +\lambda (\lambda \rho_{22}(0)-
W_{24}(\rho_{11}(0)+\lambda \rho_{33}(0)-1)) }
{\lambda  \left(W_{12} \left(\lambda
   +W_{24}\right)+\left(\lambda +W_{21}\right) \left(\lambda +W_{24}+W_{42}\right)\right)}.
\]

\end{widetext}

The fourth component $\rho_{44}$ is obtained from 
the normalization condition.
To transform back to  time domain the 
previous expressions does not comport any difficulties.
Unfortunately the results can not be cast in a compact 
form so we do not report them here. Again the off-diagonal
elements of the reduced density matrix obey the expression 
given in the main text but in in this case all the rates
$\gamma$'s are expressed via $W_{21}$ and $W_{42}$. We report the relevant ones here for convenience.

\ba
\gamma_{13}&&=(W_{21} )/2\\
\gamma_{23}&&=(W_{12}+W_{42})/2\\
\gamma_{43}&&=(W_{24})/2
\ea


\begin{thebibliography}{99}
\bibitem{qip} D.~Deutsch, Proc. R. Soc. London, Ser A {\bf 400}, 97 (1985); D.P. DiVincenzo, Science {\bf 270}, 255 (1995); A.~Ekert, and R.~Josza, \rmp {\bf 68}, 733 (1996); M.A.~Nielsen and I.L.~Chuang, {\em Quantum Computation and Quantum Information} (CUP, Cambridge, UK, 2000).
\bibitem{zurek} W.H.~Zurek, Phys. Today {\bf 44}, No. 10, 36–44 (1991); W.H.~Zurek, S.~Habib, and J.P.~Paz, Phys. Rev. Lett. {\bf 70}, 1187 (1993); W.~Zurek, Phys. Rev. D {\bf 26}, (1982)
\bibitem{joos} D. Giulini, E. Joos, C. Kiefer, J. Kupsch, I. O. Stamatescu, and H. D. Zeh, {\em Decoherence and the Appearance of the Classical World} (Springer, Berlin,1996).
\bibitem{decoherence} P.W.~Shor, W.H.~Zurek, I.L.~Chuang, and R.~Laflamme, Science {\bf 270}, 1633 (1995).
\bibitem{wot} W.K.~Wootters, \prl {\bf 80}, 2245 (1998).
\bibitem{blum} K.~Blum, {\em Density Matrix Theory and Applications} (New York: Plenum Press, 1996).
\bibitem{braun} D.~Braun, F.~Haake, and W.T.~Strunz, \prl {\bf 86}, 2913 (2001); D.~Braun, \prl {\bf 89}, 277901 (2002)
\bibitem{benatti} F. Benatti, R. Floreanini, and M. Piani, Phys. Rev. Lett. {\bf 91} 070402 (2003);
 F Benatti, R Floreanini Journal of Physics A Mathematical and General {\bf 39} 2689 (2006).
\bibitem{oh} S. Oh, and J. Kim Phys. Rev. A {\bf 73} 062306 (2006).
\bibitem{zan1} P.~Zanardi, \pra {\bf 57}, 3276 (1998); P.~Zanardi, \pra {\bf 56}, 4445 (1997)
\bibitem{wilhelm} M.J.~Storcz, and F.K.~Wilhelm, \pra {\bf 67}, 042319 (2003); M.J.~Storcz et. al., \pra {\bf 72}, 052314 (2005).
\bibitem{other} M.~Dub\'{e} and P.C.~Stamp, EInt. Journ. of Mod. Phys. B {\bf 12} 1191 (1998).





\bibitem{dfs} P.~Zanardi, and M.~Rasetti, \prl {\bf 79}, 3306 (1997); D.A.~Lidar, I.L.~Chuang, and K.B~ Whaley, Phys. Rev. Lett. {\bf 81}, 2594 (1998).
\bibitem{leggett} A.J.~Leggett, {\em et al.} \rmp {\bf 59}, 1 (1987); {\em  ibid.}, {\bf 67} 725 (E) (1995).
 U.~Weiss, {\em Quantum Dissipative Systems}
 ({\em Series in Condensed Matter Physics}, vol 13) 3rd edn
 (Singapore: World Scientific 2008)
\bibitem{petruccione} I.~Sinaysky, F.~Petruccione, D.~Burgarth, arXiv:0807.0379v1




\end{thebibliography}
\end{document}